\documentclass[sn-nature,Numbered,iicol]{sn-jnl}

\usepackage{graphicx}%
\usepackage{multirow}%
\usepackage{amsmath,amssymb,amsfonts}%
\usepackage{amsthm}%
\usepackage{mathrsfs}%
\usepackage[title]{appendix}%
\usepackage{xcolor}%
\usepackage{textcomp}%
\usepackage{manyfoot}%
\usepackage{booktabs}%
\usepackage{algorithm}%
\usepackage{algorithmicx}%
\usepackage{algpseudocode}%
\usepackage{listings}%
\usepackage{lineno}

\usepackage{geometry}
\theoremstyle{thmstyleone}%
\theoremstyle{thmstyletwo}%

\theoremstyle{thmstylethree}%

\begin{document}

\title[Exceptional sensitivity near the bistable transition point of a hybrid quantum system]{Exceptional sensitivity near the bistable transition point of a hybrid quantum system}

\author*[1]{\fnm{Hanfeng} \sur{Wang}}
\email{hanfengw@mit.edu}

\author[2,3]{\fnm{Kurt} \sur{Jacobs}}

\author[2]{\fnm{Donald} \sur{Fahey}}

\author[1]{\fnm{Yong} \sur{Hu}}
\author[1]{\fnm{Dirk R.} \sur{Englund}}
\author*[1,2]{\fnm{Matthew E.} \sur{Trusheim}}\email{matthew.e.trusheim.civ@army.mil,mtrush@mit.edu}

\affil[1]{Research Laboratory of Electronics, MIT, 50 Vassar Street, Cambridge, MA 02139, USA}

\affil[2]{DEVCOM Army Research Laboratory, Adelphi, MD 20783, USA}

\affil[3]{Department of Physics, University of Massachusetts Boston, MA 02125, USA}

\abstract{
Phase transitions can dramatically alter system dynamics, unlocking new behavior and improving performance. Exceptional points (EPs), where the eigenvalues and corresponding eigenvectors of a coupled linear system coalesce, are particularly relevant for sensing applications as they can increase sensor response to external perturbations to a range of phenomena from optical phase shifts to gravitational waves. However, the coalescence of eigenstates at linear EPs amplifies noise, negating the signal-to-noise ratio (SNR) enhancement. Here, we overcome this limitation using nonlinearity, which exhibits exceptional SNR around a bistable transition point (BP).  We couple a state-of-the-art diamond quantum sensor to a nonlinear Van der Pol oscillator, forming a self-oscillating hybrid system that exhibits both a single-valued and bistable phase. The boundaries between these phases are marked by both adiabatic and deterministic non-adiabatic transitions that enable chiral state switching and state coalescence at the BP. Crucially, NV magnetometry performed near the BP exhibits a 17× enhancement in SNR, achieving a record sensitivity of 170 fT/$\sqrt{\mathrm{Hz}}$. This result surpasses the sensitivity limit of an ideal, thermally-limited electron magnetometer and resolves a long-standing debate regarding EP-like physics in advanced quantum sensing.
}



\maketitle

\section{Introduction}\label{sec1}




Phase transitions can dramatically alter system response, enabling qualitatively new behavior and quantitatively superior performance \cite{li2023exceptional}. Exceptional points (EPs), where the eigenvalues and corresponding eigenvectors of a coupled linear system coalesce, are an example phase transition with a plethora of intriguing and counterintuitive phenomena, such as non-reciprocity and directionality \cite{yin2013unidirectional}, the revival of lasing \cite{peng2014loss}, and chiral behavior via adiabatic and non-adiabatic dynamical encirclement \cite{doppler2016dynamically, nasari2022observation, liu2021dynamically, song2021plasmonic, ding2024electrically}. 
These singularities are particularly relevant for sensing applications as they can increase sensor response to external perturbations, potentially resulting in enhanced detection of a range of phenomena from optical phase shifts to gravitational waves \cite{miri2019exceptional,wang2021coherent,ergoktas2022topological,wu2024third,loughlin2024exceptional,lai2019observation,liu2020gravitational,wu2019observation}. 

However, there is a longstanding debate on whether EPs can enhance signal-to-noise ratio (SNR) and therefore provide metrological gain \cite{loughlin2024exceptional,wang2020petermann,naikoo2023multiparameter,lau2018fundamental,chen2017exceptional,hodaei2017enhanced,kononchuk2022exceptional}. Experiments have shown increased system response near the EP \cite{chen2017exceptional,hodaei2017enhanced}, but few have claimed an increased SNR \cite{kononchuk2022exceptional,xu2024single}. Theoretical arguments against the possibility of increased SNR include that eigenvalue bifurcation in passive EP systems (without gain) is not resolvable \cite{langbein2018no,chen2019sensitivity}, and that active parity-time ($\mathcal{PT}$) symmetric systems either become unstable when perturbed from the EP or are limited by enhanced fundamental noise \cite{loughlin2024exceptional,wang2020petermann,naikoo2023multiparameter,lau2018fundamental}. 


\begin{figure*}[t!]
\includegraphics[width = 1\textwidth]{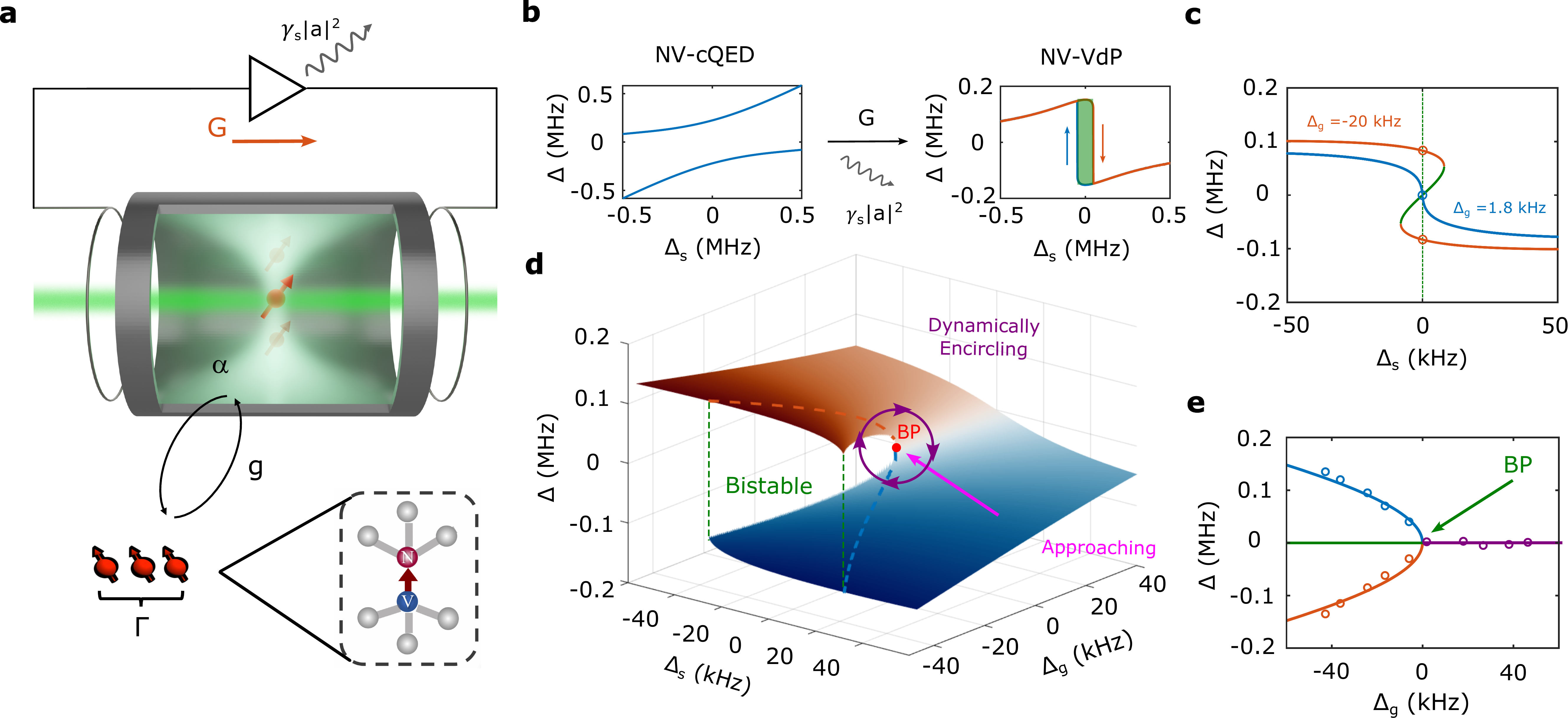}
\caption{\label{fig:1} \textbf{NV-VdP hybrid system description.} \textbf{a,} Schematic of strong coupling between an optically-polarized NV ensemble with a VdP oscillator. The VdP oscillator system is described by a cavity mode $a$ with loop gain $G$ and an amplifier nonlinear saturation parameter $\gamma_s$. The spin system has linewidth $\Gamma$ and an effective coupling strength $g$ to the cavity mode. \textbf{b,} Eigenvalue spectrum ($\Delta = \omega-\omega_c$) of a strongly-coupled NV-cavity polariton system (left) and a strongly-coupled NV-VdP system (right) using experiment parameters (see Methods). The NV-VdP system exhibits only one oscillating branch at a time, with a bistable region indicated in green. \textbf{c,} Simulated oscillating frequency. In the strong-coupling regime, three oscillation frequencies will fulfill the Barkhausen criterion (Red and green dots); two of them are stable (red), and one of them is unstable (green). Only one oscillation is achievable when $\Delta_g > 0$ (blue).  \textbf{d,} Simulated surface plot of the oscillation frequency with different $\Delta_g$ and spin detuning $\Delta_s$. Within the dashed lines, there is a bistable region where two solutions are stable. We indicate the path for dynamically encircling the BP (purple circle) and sensitivity enhancement when approaching the BP (pink line). \textbf{e,} Three solutions for $\Delta_s=0$ - two stable (red and blue) and one unstable (green) - meet at $\Delta_g=0$, forming an BP in the NV-VdP system. Dots: experimental data. Lines: simulation.}
\end{figure*}

Nonlinearity offers a potential avenue for EP-like SNR enhancement. Unlike linear EPs, nonlinear systems do not suffer from eigenspace collapse and associated compensating noise. Recent work has introduced nonlinearity to coupled electrical resonant circuits and shown that added Gaussian noise is not amplified despite an increased voltage response near the nonlinear EP, resulting in an enhanced SNR \cite{bai2024observation,bai2023nonlinearPRL,bai2023nonlinearity,suntharalingam2023noise,li2024enhanced}. It remains an open question whether this effect can be applied to leading quantum sensors to produce an SNR greater than that of non-EP operation at classical limits.
Here we couple a leading quantum sensor - an optically-polarized nitrogen-vacancy (NV) spin ensemble - to a nonlinear Van der Pol (VdP) oscillator. This system exhibits two distinct phases: a single-valued regime where only one steady-state solution exists, and a bistable regime with three potential solutions. These configurations converge at a bistable transition point (BP). Crucially, the NV-VdP has at least one real solution for all detuning and coupling parameters, enabling stable measurement of enhanced SNR around the BP and defeating prior arguments based on the instability of active EP-like sensors. The noise characteristics of the nonlinear system are also modified, resulting in converged noise around the BP that does not compensate for the signal response enhancement.

We explore several nonlinear effects around the BP in the NV-VdP hybrid system. We show that the system self-oscillates at a steady state governed by the VdP nonlinearity; exhibits bistability in the strong coupling regime that converges to single-value at the BP; shows chiral state transfer when dynamically encircling the BP; and is critically slowed at the non-adiabatic state transfer phase boundary. The demonstrations of non-adiabaticity, chirality, and slow-down are a first for a BP system and occur at a well-defined bistable phase boundary that does not exist in linear EP systems.   
We then characterize the NV-VdP system as a magnetometer near the BP. Both magnetic signal response and oscillator noise increase as the system approaches the BP, but not in proportion, leading to an SNR enhancement. We observe a 135$\times$ gyromagnetic ratio enhancement and a 17$\times$ gain in SNR, setting a record for NV sensitivity 170 $\pm$ 10 fT/$\sqrt{\mathrm{Hz}}$. Our result shows that phase-transition physics can offer a pathway to the quantum limit of sensitivity, with potential for application in a wide variety of sensor platforms.


\section{NV-VdP hybrid system}\label{sec2}

\begin{figure*}[t!]
\includegraphics[width = 1\textwidth]{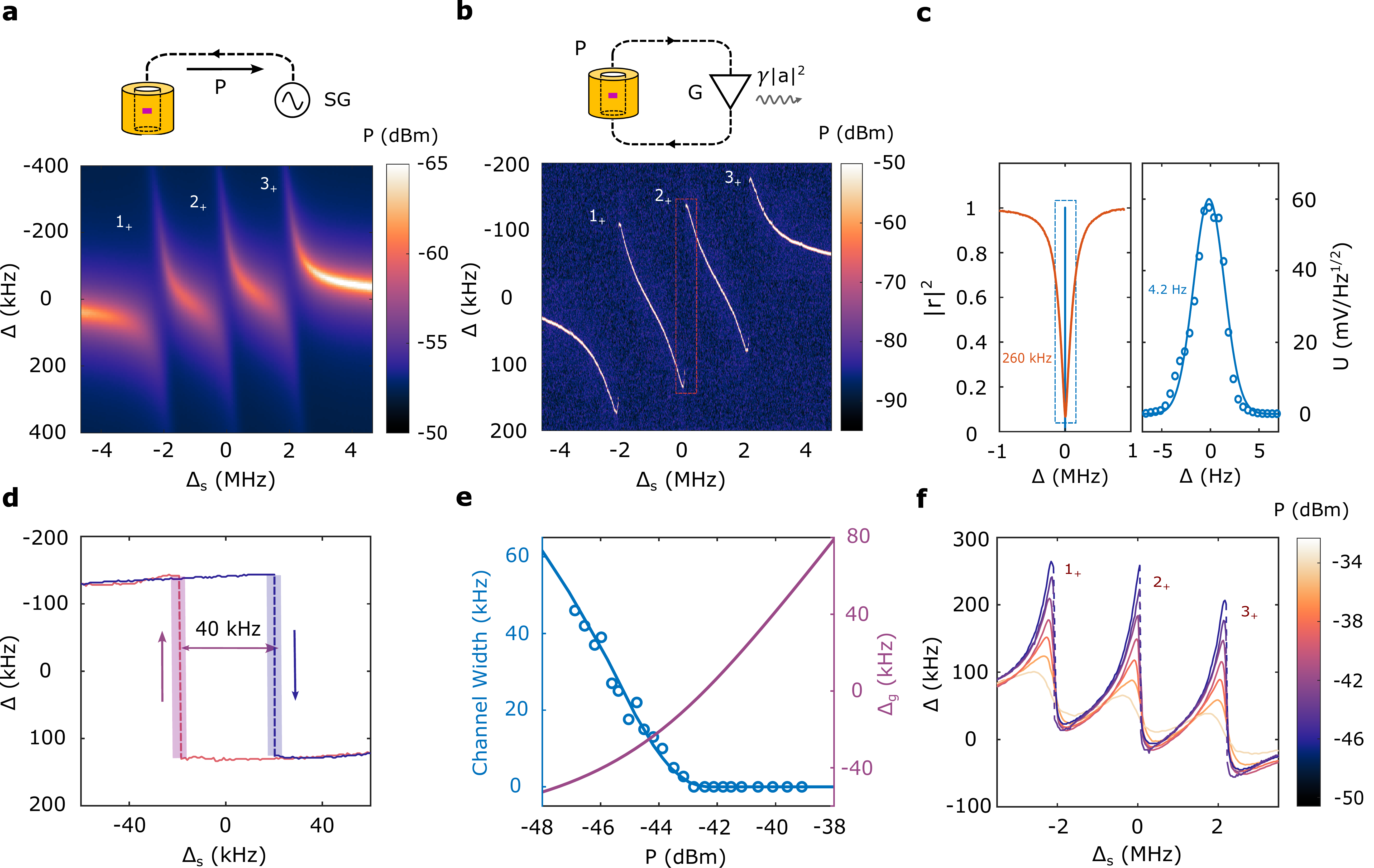}
\caption{\label{fig:2} \textbf{NV-VdP spectroscopy.} \textbf{a,} The NV-cQED reflection spectrum. Three transitions $1_+$, $2_+$, and $3_+$ shown due to the hyperfine coupling with $^{14}N$. \textbf{b,} NV-VdP self-oscillating spectrum. \textbf{c,} Comparison of reflection ($|r|^2$) and self-oscillating spectra (amplitude spectral density) at $\Delta_s = 2\pi\times 5$ MHz. We observe a self-oscillating linewidth of $2\pi\times4.2$ Hz (blue; zoom-in on right), which is 5 orders of magnitude lower than the cavity linewidth $\kappa=2\pi\times260$ kHz (red). \textbf{d,} The self-oscillation spectrum with increasing (blue) and decreasing (red) spin-cavity detuning around $\Delta_s = 0$ (red box in \textbf{b}). A hysteresis window of 40 kHz is seen, with the shaded region indicating the standard deviation of the transition frequency. \textbf{e,} Left scale: the hysteresis channel width with loop oscillator power. Right scale: Relation between $\Delta_g$ and oscillator power. The hysteresis channel width decreases with increasing $\Delta_g$ and is reduced to zero when $\Delta_g=0$. \textbf{f,} Self-oscillator center frequency as a function of spin detuning, for different oscillating power $P$. Dashed lines indicate discontinuous jumps in frequency, which disappear at higher power. the presence of non-zero phase shift in the non-cavity component of the loop can result in asymmetric Fano-type lineshapes where the oscillator frequency does not match the bare polariton resonance (see Supplementary Information Sec. I).}
\end{figure*}

The nonlinear NV-VdP system builds upon a linear system comprising an NV ensemble strongly-coupled to a microwave cavity mode (Fig. 1a, center). NV systems have been widely investigated as leading solid-state quantum magnetometers \cite{barry2020sensitivity,du2017control,degen2017quantum,boss2017quantum,schmitt2017submillihertz}, and coupled NV-cavity devices have enabled sub-pT sensitivities for magnetometry \cite{wang2024spin}. The Hamiltonian describing the cavity quantum electrodynamic (cQED) system with the Holstein-Primakoff transformation is \cite{zhang2021exceptional}:
\begin{equation}
    H_{\mathrm{eff}} = \begin{pmatrix}
    \omega_c - i\kappa/2 & g\\
    g & \omega_s - i\Gamma/2
    \end{pmatrix}
\end{equation}
Here $\omega_c$ is the cavity frequency, $\omega_s$ is the spin ensemble center frequency, $\kappa$ is the cavity loss rate, $\Gamma$ is the spin linewidth, and $g$ is the coupling strength. 

\subsection{Nonlinear NV-VdP hybrid system}

We form an active, nonlinear system by placing the linear microwave cavity in a feedback loop, creating a VdP oscillator. Adding gain and nonlinearity modifies the effective cavity frequency to $\omega_c + i(G-\kappa/2-\gamma_s|a|^2)$ \cite{kanamaru2007van,yao2023coherent}. Here $G$ is the one-photon gain, $\gamma_s$ is the two-photon nonlinear damping rate corresponding to amplifier saturation, and $a$ is the cavity field operator. When the loop gain balances loss, the VdP system self-oscillates and produces a coherent monochromatic output tone. Unlike in the bare NV-cQED system where multiple polaritons appear simultaneously near $\Delta_s = 0$, the NV-VdP oscillation only produces a single output tone for all parameters due to gain competition, enabling clear state identification shown in Fig. 1b.

Solving the coupled-mode equations gives a cubic equation for the steady-state frequency. Near resonance in the strong-coupling regime $\Delta_g = \Gamma/2 - g < 0$, there are three real-valued solutions for the oscillation frequency, as shown in Fig. 1c. Two of these are stable to perturbations while the third is not observable in the steady state \cite{angerer2017ultralong}. This is indicative of bistability, which disappears non-adiabatically as the spins are detuned from the cavity resonance, as shown in Fig. 1d. As the coupling strength is reduced, the range of spin-oscillator detuning for which the system is bistable (hysteresis window) is also reduced, eventually to zero. Beyond this point ($\Delta_g >0$) there is only a single real-valued solution for the oscillation frequency, and we denote the point at which the bistability vanishes as the BP ($\Delta_g =0$), see Fig. 1e. There is always a stable steady-state solution for the NV-VdP above the critical gain threshold, even when detuned or near the BP, in comparison to a standard EP where small deviations can lead to complex (lossy or diverging) solutions.

\subsection{NV-VdP characterization}

We first characterize the passive NV-cQED system without loop gain. The reflection spectrum is shown in Fig.\ 2a. The clear avoid-crossing indicates high cooperativity $C \sim 2$, and $g > \max\{\kappa/2, \Gamma/2\}$ shows that the NV-cavity system produces a hybrid polariton in the strong-coupling regime. Three resolved transitions correspond to the three $^{14}N$ hyperfine subensembles, separated by $A_{zz} = 2.1$ MHz.

We then apply a loop gain using amplifiers and attenuators in series to achieve a low and tunable steady-state power (see Methods). As the amplifier gain surpasses the loop losses, the system oscillates from an initial thermal population. The resulting self-oscillator output spectrum is shown in Fig.\ 2b. Only a single oscillating mode appears, and its frequency exhibits a discontinuous jump when tuning the spins across resonance.

The feedback gain results in a significant reduction in the effective linewidth, as shown in Fig.\ 2c. The VdP linewidth (4 Hz) is 5 orders of magnitude smaller than the cavity linewidth ($\kappa = 260$ kHz). The equivalent quality factor of this oscillator, $Q \sim 10^9$, significantly surpasses that of microwave resonators operated without feedback. 

A key property of polaritonic self-oscillators is a hysteresis arising from the bright-state selection (gain competition) \cite{yao2023coherent}, which delineates the bistable regime. In Fig.\ 2d, we plot the steady-state output frequency of the self-oscillator as a function of increasing (blue) and decreasing (purple) $\Delta_s$. In this setting, the gain remains active while adjusting the spin frequency, so the initial condition for each subsequent detuning is the steady-state value from the previous one. A polariton branch-switching discontinuity occurs at $\Delta_s=\pm 20$ kHz for upward and downward sweeps.

To tune the system to the BP, we vary the oscillating power by changing the loop attenuation. As the loop power increases, the spin saturation increases in turn, decreasing the total spin-cavity coupling strength \cite{wang2024spin}. This behavior is shown experimentally and modeled by combining the nonlinear spin saturation with the self-oscillating model (Fig. 2e). The hysteresis channel width reduces to zero at a $P=-42.5$ dBm, corresponding to the disappearance of the frequency-jump discontinuity (Fig.\ 2f) and the formation of the BP.

\begin{figure*}
\includegraphics[width = 1\textwidth]{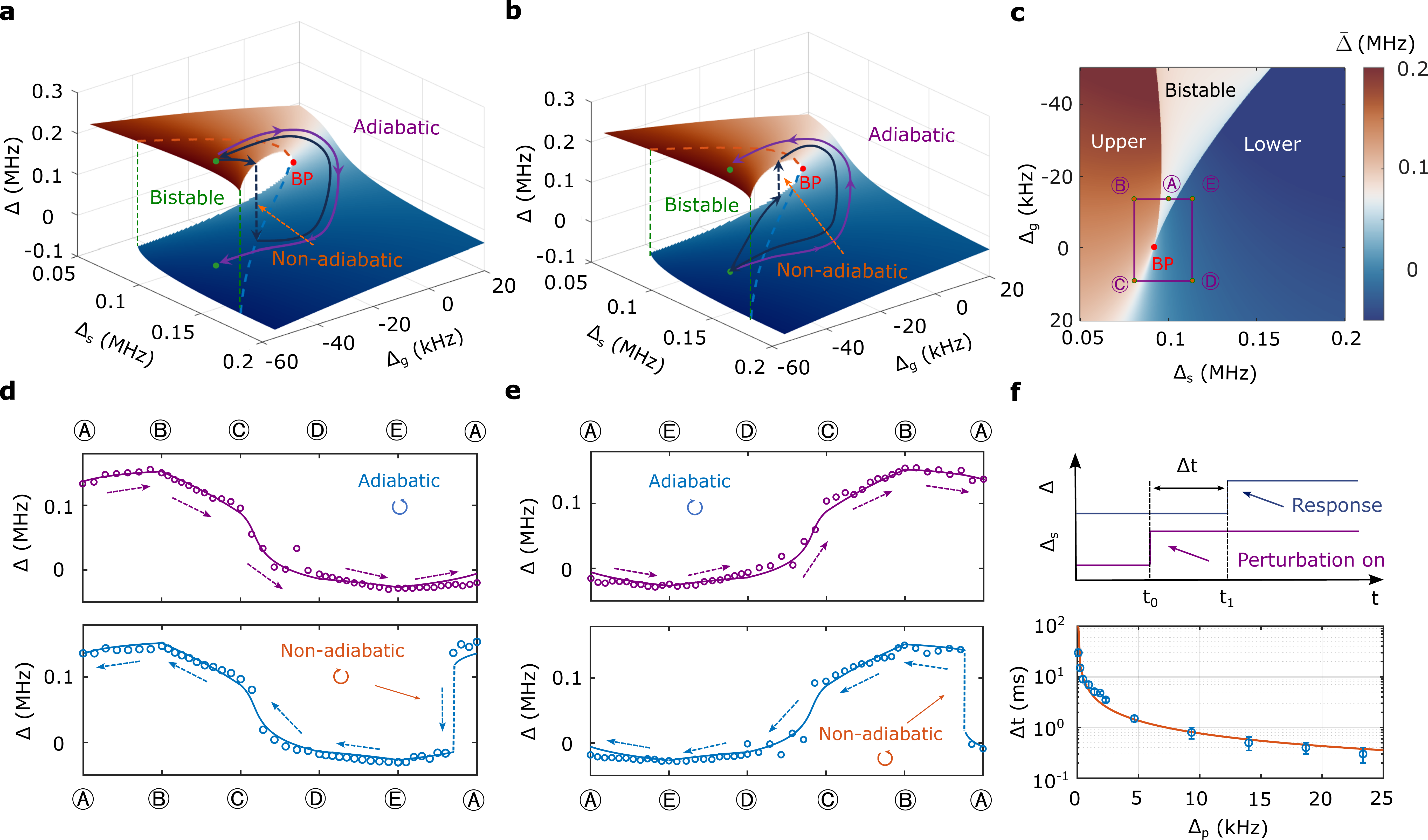}
\caption{\label{fig:3} \textbf{Dynamically encircling the BP.} \textbf{a,} Encircling the BP adiabatically (clockwise) and non-adiabatically (counterclockwise) starting from the lower branch in the bistable regime. Two trajectories end in different states. \textbf{b,} Same encircling trajectories with \textbf{c,} but the starting point is on the lower branch. \textbf{c,} Dynamical encirclement trajectory in the 2D parameter space. The color axis indicates the average frequency of the stable branches for that parameter combination $\bar{\Delta} = (\Delta_1+\Delta_2)/2$. Note that the bistable region is asymmetric as changing $\Delta_g$ induces a variable loop phase which is not re-zeroed in this measurement.
\textbf{d,} The experimental data and theory plot for the adiabatic (non-adiabatic) encircling A-B-C-D-E-A (A-E-D-C-B-A) starting from the upper branch. \textbf{e,} The experimental data and theory plot for the adiabatic (non-adiabatic) encircling A-E-D-C-B-A (A-B-C-D-E-A) starting from the lower branch. \textbf{f,} Transition-edge measurement. Top: Measurement schematic. A variable spin frequency change $\Delta_p$ is turned on at $t_0$, shifting the spins across the transition boundary, and an oscillator state change is observed at $t_1$. The time delay between these events is $\Delta t = t_1-t_0$. Bottom: Measured $\Delta t$ as a function of step size $\Delta_p$ (blue points) and stretched exponential fit (red curve).}
\end{figure*}

\section{EP-like phenomena at the BP}

\subsection{Chirality and Dynamical Encirclement}

One intriguing behavior shown in EPs is chirality in state change dynamics along different parameter trajectories \cite{doppler2016dynamically, nasari2022observation, liu2021dynamically, song2021plasmonic, ding2024electrically}. This arises from the non-orthogonality of the linear eigenstates, which can result in a non-adiabatic flipping process between eigenstates. As the state non-orthogonality is associated with the non-Hermetian loss processes, encirclement is highly lossy in linear EPs, or requires ``hopping" with fast Hamiltonian parameter changes\cite{li2023exceptional}. In the BP system, the oscillator states are well-defined and orthogonal and  the non-adiabatic flip occurs deterministically when crossing the phase boundary. Here we demonstrate these chiral processes around the BP using the NV-VdP system.


We consider $(\Delta_g,\Delta_s$) parameter-space trajectories that encircle the BP starting from within the bistable region, as shown in Fig. 3a-c. We start with an initial state on either the upper (Fig.\ 3a) or lower (Fig.\ 3b) branch of the oscillator and proceed with either a ``clockwise" or ``counterclockwise" loop. The parameter trajectory forms a loop through points A-E in the parameter space, as shown in Fig.\ 3c, differing in path order and initial state. 

The resulting behavior is shown in Figs.\ 3d,e. We start from the upper branch within the bistable regime (A). Fig.\ 3d top shows adiabatic encircling the BP by A-B-C-D-E-A. This encircling the BP flips the oscillator state from the upper to lower branch adiabatically. On the other hand, when encircling the BP counterclockwise along the path A-E-D-C-B-A (Fig.\ 3d bottom), the system undergoes a non-adiabatic transition between A-E, and the final state is preserved. Thus the system exhibits chiral behavior, where the change in oscillator state depends on the directionality of the encircling. Analogous behavior is shown in Fig.\ 3e in the case starting from the lower branch. Our result illustrate the first time this distinction between the BP and linear EP systems, where the persistence of a stable solution guarantees robust and deterministic chiral state transfer without extensive loss or Hamiltonian hopping \cite{doppler2016dynamically}. 


While the chiral state trajectories are adiabatic, transitions across the hysteresis boundary (e.g. A$\rightarrow$E) can result in sudden state change. These effects are properly termed ``non-adiabatic" as the transition time approaches infinity as the detuning step becomes infinitely small. This scaling appears in the time-domain response near the phase boundary (Fig. 3f). As the hybrid oscillator is perturbed across the transition boundary with an increasingly small step $\Delta_p$, the time for the system to switch states scales with $\Delta t\propto |\Delta_p|^\xi$ with $\xi= -0.83$. Thus, adiabatic passage is not possible across this boundary, in contrast to the trajectories along connected Riemann surfaces (see Supplementary Information Sec. III).

\subsection{SNR enhancement in BP magnetometry}

In this section, we demonstrate BP-enhanced magnetometry with the NV-VdP. Self-oscillating magnetometers have achieved high performance without requiring an external probe tone \cite{liu2020noise,schwindt2004chip,jenkins2013self,barry2023ferrimagnetic}. The performance of a general self-oscillating magnetometer can be calculated using the oscillator phase noise and the gyromagnetic ratio of the magnetically sensitive component \cite{barry2023ferrimagnetic}. For an electronic spin system coupled to a Leeson's oscillator with finite output power, there is a resulting lower bound on the magnetic sensitivity given by $\sim \sqrt{3/2}f_L\sqrt{k_BT/P}/\gamma_e$. Using our system parameters of Leeson's frequency $f_L = 155$ kHz,  $P \sim -40$ dBm, and the NV gyromagnetic ratio $\gamma_e = 28$ GHz/T, a simple analysis would predict a sensitivity no better than 2 pT/$\sqrt{\mathrm{Hz}}$. The limitation for a linear sensor is similar for equivalent power and cQED parameters (See Supplementary Information Sec. V). Here we show that our hybrid self-oscillator can break this limit by enhancing the effective gyromagnetic ratio near the BP.

\begin{figure*}
\includegraphics[width = 1\textwidth]{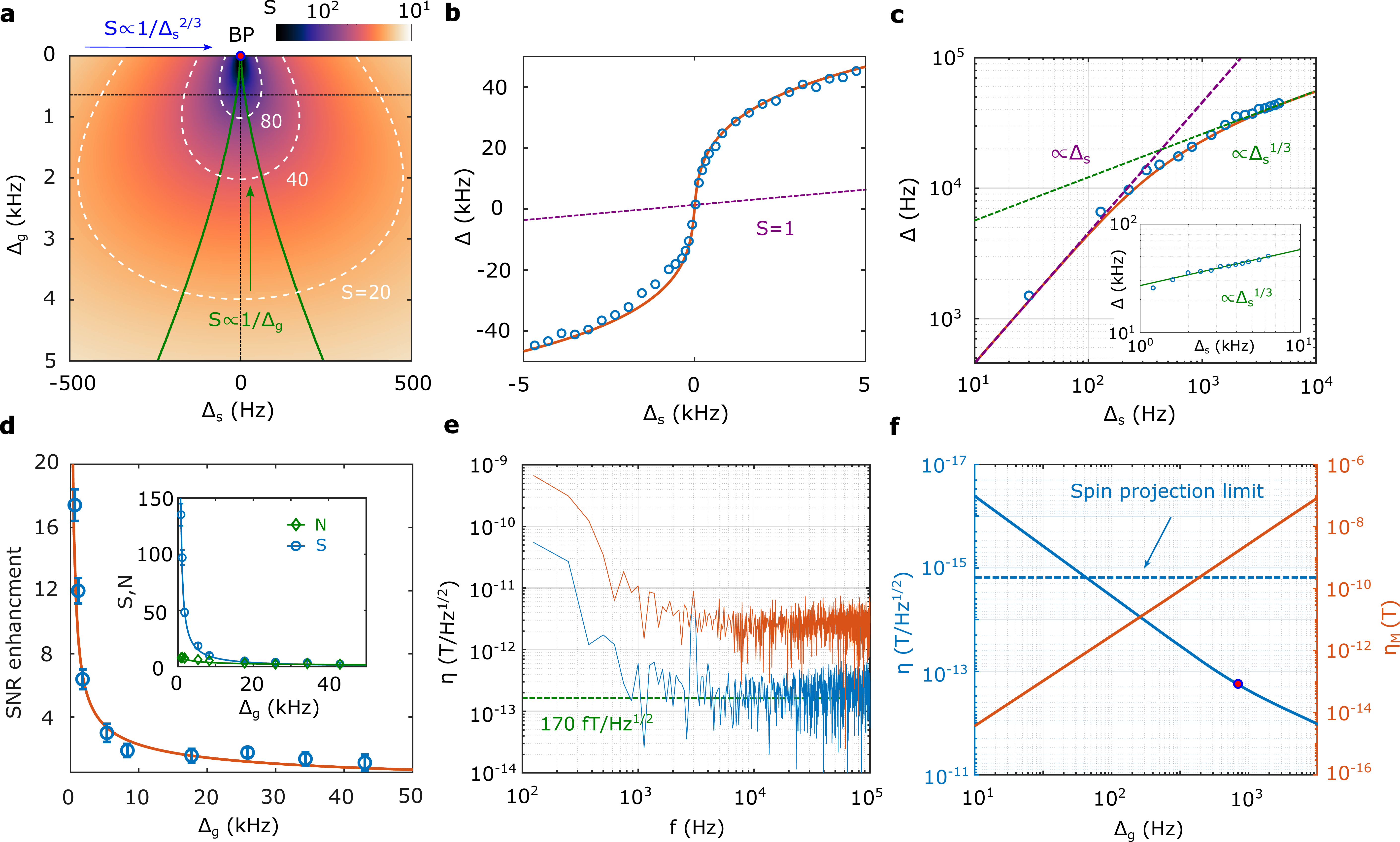}
\caption{\label{fig:4} \textbf{BP-enhanced magnetometry.} \textbf{a,} Modeled response enhancement $S$ as a function of $\Delta_s$ and $\Delta_g$. The green region shows the linear response region where $S$ is approximately constant (within 1\%). The black lines are the trajectories shown in \textbf{b-d}, and the white lines are contours of constant $S$. \textbf{b,} The NV-VdP frequency $\Delta$ as a function of spin detuning $\Delta_s$ at $\Delta_g=0.6$ kHz. Blue points are experimental data, the orange curve is the calculated frequency from numerical theory, and the purple dashed line indicates linear response at a bare electron $\gamma_e$. The system shows a maximum response $S_{\mathrm{max}} = 133$. \textbf{c,} Logarithmic plot of NV-VdP response with $\Delta_g=1.8$ kHz. The sensor response can be described by $\Delta\propto\Delta_s^{1/3}$ for $\Delta_s > 1$ kHz (green line, inset figure), but remains linear for small $\Delta_s$ (purple line). \textbf{d,} SNR as a function of $\Delta_g$ around zero spin detuning. We achieve a 17$\times$ enhancement at $\Delta_g=0.6$ kHz. Inset: The responsivity (blue) and noise enhancement (green) for different $\Delta_g$. \textbf{e,} Magnetic field sensitivity for a measurement time of 8 ms with $\Delta_g=0.6$ kHz (blue). The mean sensitivity in a 1 kHz band around 10 kHz offset is 170 $\pm$ 10 fT/$\mathrm{\sqrt{Hz}}$. We plot $\Delta_g=46.1$ kHz as a comparison (red). \textbf{f,} Sensitivity prediction assuming a Leeson's oscillator phase noise profile with power $P=-40.5$ dBm. When $\Delta_g$ approaches 0, the sensitivity improves (blue) but the linear response range also decreases (red). Note that we flipped the left $y$-axis to show the improved sensitivity is traded-off with the smaller dynamical range. Red dot: our experiment point.}
\end{figure*}

We first characterize the oscillator's responsivity around the BP. We define a relative response, or effective gyromagnetic ratio enhancement, as the ratio between the change in oscillator frequency and spin frequency $S = d\Delta/d\Delta_s$. We plot the modeled value of $S$ as a function of $\Delta_s$ and $\Delta_g$ in Fig. 4a. When $\Delta_g=0$, the scaling obeys $S\propto1/\Delta_s^{2/3}$ ($\Delta\propto\Delta_s^{1/3}$), corresponding to a third-order BP due to the VdP nonlinearity. When $\Delta_g>0$, $S$ is nearly constant in the regime $|\Delta_s|\ll \Delta_g$ resulting in enhanced but linear response. In this regime, the maximum signal response scales $S_{\mathrm{max}}\propto 1/\Delta_g$, faster than the scaling $S_{\mathrm{max}}\propto 1/\sqrt{\Delta_g}$ seen in an EP formed by two coupled linear systems (see Supplementary Information Sec. I). 

To demonstrate the predicted scaling experimentally, we measure the self-oscillation frequency while changing the spin detuning using a bias magnetic field $\Delta_s = \gamma_eB$ for a fixed coupling offset $\Delta_g=0.6$ kHz with a maximum $S_{\mathrm{max}} = 133$ (Fig.\ 4b). The oscillation frequency scales as $\Delta\propto\Delta_s$ when $\Delta_s < 250$ Hz, and continuously changes to $\Delta\propto\Delta_s^{1/3}$ when $\Delta_s>\Delta_g$, as shown in Fig. 4c. 


We then use a test field to demonstrate the SNR enhancement. We bias the system around $\Delta_s=0$ for different $\Delta_g$ while applying a test field of $0.14$ nT at 3 kHz. We calculate the effective gyromagnetic ratio of the self-oscillator as $\gamma_\mathrm{eff} = \sqrt{3}/B_{t}\frac{d\phi}{dt}$ where $B_t$ is the test field amplitude, and simultaneously measure the relative noise amplitude $N$ at 30 kHz offset. The signal and noise enhancement factors $S$ and $N$ are the ratio of the observed gyromagnetic ratio and noise amplitude to non-BP reference values of the bare electron gyromagnetic ratio and spin-detuned noise, respectively. The resulting SNR is shown in Fig. 4d, with signal $S$ and noise $N$ shown independently in the inset.


The signal and noise enhancements have different scalings when approaching the BP. As $\Delta_g$ decreases the AC signal enhancement $S$ increases $S\propto 1/\Delta_g$, reaching a maximum enhancement of 135 $\pm$ 9 at $\Delta_g=0.6$ kHz. This matches the theoretical modeling and the response enhancement observed in the quasi-DC bias field sweep experiment.  

Unlike the diverging signal enhancement, the noise enhancement:
\begin{equation}
   S_{pp}(\Delta_g,\Delta_s=0)\propto \frac{(g+\Delta_g)^2}{(g+\Delta_g)(g+\gamma_s|\alpha|^2)-g^2}
\end{equation}
does not diverge when approaching the BP (see Supplementary Information Sec. II). This model reduces to the linear EP case when the nonlinear component $\gamma_s|\alpha|^2\rightarrow 0$, leading to the diverging noise $S_{pp}\propto 1/\Delta_g$ predicted in the linear EP systems \cite{loughlin2024exceptional}. In our system, however, the nonlinear term gives a nonzero offset for the denominator of Eq. (2), preventing the noise from diverging. When $\Delta_g\rightarrow 0$, Eq. (2) can be written as:
\begin{equation}
   S_{pp}(\Delta_g\rightarrow 0)\propto \frac{g}{\gamma_s|\alpha|^2}
\end{equation}
which is independent with either $\Delta_s$ or $\Delta_g$. For a fixed $\alpha$ due to the spin saturation, a larger $\gamma_s$ will give a better noise performance. While the noise is not diverging, it is enhanced, with an observed $N = 8$ at the closest approach to the BP ($\Delta_g=0.6$ kHz). 

The magnetic response improvement and increased noise do not negate each other, finally resulting in a $17\times$ sensitivity (SNR) enhancement near the BP (Fig. 4d). The frequency-domain magnetic noise floor at $\Delta_g=0.6$ kHz is shown in Fig. 4e.  We achieve a record sensitivity of 170 $\pm$ 10 fT/$\sqrt{\mathrm{Hz}}$ around 10 kHz. This BP-enhanced performance breaks the sensitivity bound for a general thermally-limited, bare-electron device by a factor of 10 at equal oscillating power. 
These results establish the state-of-the-art NV sensitivity across all readout and control methodologies \cite{barry2020sensitivity,wang2024spin,barry2024sensitive} and have a further advantage of not requiring an external signal generator. 



The ultimate achievable sensitivity depends both on the sensitivity of the underlying NV-cQED system \cite{wang2024spin} and the BP enhancement. BP enhancement comes with a cost in the system dynamic range, as shown in Fig. 4f. Our device is in an unshielded laboratory environment, limiting the precision of BP approach and therefore the current NV-VdP performance. 
Improving the NV-cQED system can potentially improve both bare sensitivity and BP enhancement, offering several orders of magnitude improvement (see Supplementary Information Sec. VI).
Our semiclassical model does not capture quantum noise, and predicts a diverging sensitivity when closely approaching the BP. The spin-projection limit is shown in Fig. 4f, and the role of spin quantization effects near $\Delta_g=0$ are an area for future investigation.

\section{Discussion and Outlook}

The NV-VdP platform offers several directions that push the envelope of physics. Operation of the device near the critical point (gain threshold), or dynamically into the deep spin-saturation regime (frequency/power coupling) offer potential new phases and nonlinear phenomena. The current device can be modeled using a semi-classical description as it does not exhibit a non-classical (entangled or squeezed) output and the large number of spins and photons are well-approximated by a harmonic oscillator. The NV system, however, can be cooled to its quantum mechanical ground state even in ambient temperatures \cite{fahey2023steady,zhang2022microwave,wu2021bench}, which paves a path towards a strongly-coupled NV-VdP system in the quantum regime. Closer approach to the BP also challenges quantum limits as SNR continuous to increase. Operation at the standard quantum limit of spin-projection noise would result in an additional enhancement of $\sim$ 100, while reaching the Schawlow-Townes limit of microwave phase noise would result in an enhancement of $ k_BT/h\nu \sim$ 1000. Future devices harnessing topological and phase-change behavior have the potential to become standard in highly sensitive magnetometry and other areas of quantum sensing, as well as testbeds for nonlinear phenomena and phase transition physics.

\bibliography{sn-bibliography}

\section*{Methods}

\subsection*{Experiment setup}

The setup is shown in Extended Data Fig. 1, and can be treated in two parts: the microwave circuitry that forms the loop oscillator, and an optically pumped diamond-cavity interface. The diamond-cavity system is similar to our previous work \cite{wang2024spin}, with the primary difference being the addition of a second coupling loop to allow transmission through the NV-cavity system. As in previous, the diamond (3 mm $\times$ 3 mm $\times$ 0.9 mm, 4 ppm NV ensemble, sourced from Element 6) is set at the TE01$\delta$ mode maximum point in the center of the dielectric resonator (Skyworks, $\varepsilon_\mathrm{r} \sim 31$, Material: D-87XX). A wafer of 4H-SiC is used for heat transfer and supporting the diamond, while two pieces of low-loss-tangent polytetrafluoroethylene (PTFE) are used to fix and align the dielectric resonator. An aluminum shield is employed to isolate the system from external lab signals, such as WiFi and 3G signals (1.9 GHz), and to reduce radiative losses. An 8W 532 nm pump laser (Sprout-G) optically polarizes the spin ensemble. Three-axis magnetic coils (not shown) provide an external magnetic bias field. 

The microwave circuitry contains an amplifier chain ``Amplifier 1 (RLNA02G04G60) -- 30 dB attenuation -- Amplifier 2 (RLNA02G08G30) -- 3 dB attenuation -- Amplifier 3 (RLNA02G08G30S) -- 20 dB attenuation -- tunable attenuation" with a 10:1 directional coupler (PE2CP1102) to guide a part of the signals outside of the self-oscillation loop for measurement. The series of low noise amplifiers and attenuators ensures that the loop gain is always greater than loss, but allows for tuning of the steady-state oscillating power. A spectrum analyzer and a custom heterodyne receiver are used to analyze the output tone from the self-oscillation system. The signal from the directional coupler is mixed down to an intermediate frequency with an external signal generator (SMA100B) through a mixer (HX3400) and then digitized by a DAQ.

\subsection*{Hilbert transform}

We use the Hilbert transform to recover the magnetic field \cite{barry2023ferrimagnetic} from the heterodyne measurement. This process isolates the instantaneous phase $\phi(t)$ from variations in the instantaneous amplitude. Two conditions need to be met for the Hilbert transform to be valid: the additive phase noise must be small, and the additive phase noise and amplitude noise must be slowly varying ($\sim$ 3 kHz in our experiment) compared with the intermediate frequency of the mixer ($\omega_I\sim$ 0.5 MHz in our experiment). As explained above, both requirements are fulfilled in our experiments. 

The data we get from the experiment can be expressed as a real-valued waveform:
\begin{equation}
    v(t)=V_0(1+\alpha(t))\cos(\omega_i t+\varphi(t))
\end{equation}
Given $|\varphi(t)| \ll 1$, we can write the equation as:
\begin{equation}
    v(t) \approx V_0[1+\alpha(t)]\left[\cos \left[\omega_i t\right]-\varphi(t) \sin \left[\omega_i t\right]\right]
\end{equation}
Denoting the Hilbert transform of $v(t)$ as $\hat{v}(t)$, we have:
\begin{equation}
    \hat{v}(t) \approx V_0[1+\alpha(t)]\left[\sin \left[\omega_i t\right]+\varphi(t) \cos \left[\omega_i t\right]\right]
\end{equation}
Using single approximations, we finally obtain the resulting signal:
\begin{equation}
    v(t)+i \hat{v}(t) \approx V_0[1+\alpha(t)] e^{i\left(\omega_i t+\varphi(t)\right)}
\end{equation}
We now observe that the amplitude noise and phase noise are completely separated. The time domain magnetic field waveform $B(t)$ can then be determined by:
\begin{equation}
    B(t)=\frac{1}{\gamma} \frac{d \varphi(t)}{d t}
\end{equation}


\subsection*{BP magnetometry}

To obtain the responsivity in main text Fig. 4b,c, we applied a triangular wave with a period of 200 ms and an amplitude of 0.6 $\mu$T and used the heterodyne circuit to detect the real-time voltage $V(t)$. We segment the real-time data with a window size depending on the desired resolution and define the oscillator frequency as the frequency with the highest power in each window. For each period of the triangle wave, we fit the frequency data to the quantum Heisenberg model with a spin-frequency offset as a free parameter. This allows us to correct for drifts in magnetic field over time due to low-frequency noise sources. We then re-center each period, average the time-series frequency data, and re-fit to the quantum Heisenberg model to extract the underlying coupling strengths and linewidths and produce a fitting maximum responsivity. The final fitted model is shown alongside the averaged (drift-corrected) data in the main text Fig. 4b,c. 

Given a specific period $p$, there will be a tradeoff between the window size and the resolution for $\Delta_s$. Assuming the window size $w$, the oscillating frequency accuracy is $1/w$. The minimum $\Delta_s$ is then $1/Sw$ with $S$ the responsivity enhancement. Ideally, we want to cover the linear regime for our measurement so the magnetic field is lower-bounded by $B=\Delta_g/\gamma_e$. Then the resolution for $\Delta_s$ is limited by $\Delta_g w/p$. Therefore the oscillating frequency accuracy is traded off with the spin detuning frequency resolution. In our measurement, the window size was set to be 0.5 ms, which resulted in an oscillator frequency resolution of 2 kHz and $\Delta_s$ resolution of 98 Hz for a single measurement. The total measurement duration in Fig. 4 was 2 seconds or 20 total sweeps across the BP.

To determine the magnetic sensitivity, we set the bias magnetic field at the point of maximum slope determined by the triangular wave measurement and measured the output voltage for 0.5 s at each microwave power. We then segmented each 0.5-second dataset with a window size of 8 ms and calculated the PSD individually for each window with the Hilbert transform. The voltage was converted to the magnetic field with the calibration by the test field at 3 kHz: we first calibrated a magnetic field of 0.14 nT (RMS value) using the cQED sensor functioning in the linear regime (the calibration process is described below). We then used this calibrated magnetic field to convert the voltage amplitude spectral density (unit: V/$\sqrt{\mathrm{Hz}}$) to the magnetic field ASD (unit: T/$\sqrt{\mathrm{Hz}}$). We subsequently define the sensitivity as the magnetic amplitude noise floor at within a 125 Hz bandwidth around 30 kHz (equivalent to the ratio of the voltage amplitude in the test field band to the voltage noise amplitude at 30 kHz). This process was repeated for each microwave power to generate Fig. 4 in the main text. 

Magnetic field calibration process: The cQED sensor used to calibrate the magnetic field was calibrated by the reference magnetometer by an existing magnetometer: 
We used a coil with turn number $N= 400$, radius $r=7$ cm, and distance between the coil and sensor $d=24$ cm. We use a DC power supply to generate a voltage $U=5$ V, resulting in a current output of $I=216$ mA. The magnetic field generated by this test coil at the sensor position is $B_{\mathrm{cal}}=2\mu_0\pi r^2NI/4\pi(d^2+r^2)^{3/2} = 0.136~G$, which is similar with the measurement from the measurement from a commercial gaussmeter $B_{\mathrm{Gauss}} = 0.13(1)~  G$. The sensor shows a response of $B_{\mathrm{cQED}} = 0.134~G$ with a 2-3\% error compared with the reference magnetometer and calculated result. The difference can be attributed to the estimation of the size parameters and the twisted angle of the coil with the diamond surface.

\subsection*{Extracting experimental parameters}

In Fig. 2 and Fig. 3, we fit the linear spectrum and hysteresis window to get the parameters: $\kappa = 2\pi\times 320$ kHz, $\kappa_{c1}=2\pi\times 130$ kHz, $\kappa_{c2} = 2\pi\times 50$ kHz, individual spin coupling strength $g_s = 2\pi\times 0.019$ Hz, NV homogeneous linewidth $\gamma = 2\pi\times 35$ kHz, coupling strength $g = 2\pi\times 0.22$ MHz, and NV linewidth $\Gamma_1 = 2\pi\times 0.315$ MHz. Those parameters are then used in Fig. 1 for schematic descriptions. In Fig. 4, the best fitting shows the parameters: $\kappa = 2\pi\times 320$ kHz, $\kappa_{c1}=2\pi\times 130$ kHz, $\kappa_{c2} = 2\pi\times 50$ kHz, individual spin coupling strength $g_s = 2\pi\times 0.019$ Hz, NV homogeneous linewidth $\gamma = 2\pi\times 35$ kHz, coupling strength $g = 2\pi\times 0.22$ MHz, and NV linewidth $\Gamma_1 = 2\pi\times 0.301$ MHz. The slight difference in the coupling strength and total spin linewidth can be attributed to laser alignment and bias magnetic field alignment.  


In Fig. 1 and Fig. 4, we define $\Delta_g=\Gamma/2-g$ to illustrate the behavior of approaching the BP. In general, there is a phase delay when changing the tunable attenuation. We reset the loop phase to zero, and the BP will arise at $\Delta_g=0$. In Fig. 2 and Fig. 3, we add the loop phase and the BP will not arise at $\Delta_g=0$. To make the discussion consistent, we define $\Delta_g=\Gamma/2-g-\xi$ with $\xi$ an offset. In Fig. 2 and Fig. 3, $\xi$ is set to be 2 kHz.

\section*{Acknowledgments\protect}
The authors would like to thank Reginald Wilcox and Shuang Wu for their helpful discussions. H.W.~acknowledges support from Bosch Inc. D.R.E.~acknowledges funding from the MITRE Corporation and the U.S.~NSF Center for Ultracold Atoms. 

\section*{Author Contribution}
H.W. and M.E.T. created the setup and conducted the experiments. H.W., K.J., and M.E.T. developed the theory. D.F. and M.E.T. conducted the preliminary experiment. 
H.W. and M.E.T. prepared the manuscript. Y.H. helped with the time domain measurements.
All authors discussed the results and revised the manuscript. M.E.T. and D.R.E. supervised the project. 

\section*{Data availability} The data and code that support the findings of this study are available from the corresponding author upon request.


\end{document}